\documentstyle[11pt]{article}
\begin{document}
\ifx\TwoupWrites\UnDeFiNeD\else\target{\magstepminus1}{11.3in}{8.27in}
	\source{\magstep0}{7.5in}{11.69in}\fi
\newfont{\fourteencp}{cmcsc10 scaled\magstep2}
\newfont{\titlefont}{cmbx10 scaled\magstep2}
\newfont{\authorfont}{cmcsc10 scaled\magstep1}
\newfont{\fourteenmib}{cmmib10 scaled\magstep2}
	\skewchar\fourteenmib='177
\newfont{\elevenmib}{cmmib10 scaled\magstephalf}
	\skewchar\elevenmib='177
\newif\ifpUbblock  \pUbblocktrue
\newcommand\nopubblock{\pUbblockfalse}
\newcommand\topspace{\hrule height 0pt depth 0pt \vskip}
\newcommand\pUbblock{\begingroup \tabskip=\hsize minus \hsize
	\baselineskip=1.5\ht\strutbox \topspace-2\baselineskip
	\halign to\hsize{\strut ##\hfil\tabskip=0pt\crcr
	\the\Pubnum\crcr\the\date\crcr}\endgroup}
\newcommand\YITPmark{\hbox{\fourteenmib Last\hskip0.2cm
Number\hskip0.2cm from \hskip0.2cm YITP\hskip0.2cm
    \fourteenmib Uji\hskip0.15cm Research\hskip0.15cm Center\hfill}}
\renewcommand\titlepage{\ifx\TwoupWrites\UnDeFiNeD\null\vspace{-1.7cm}\fi
	\YITPmark
\vskip0.6cm
	\ifpUbblock\pUbblock \else\hrule height 0pt \relax \fi}
\newtoks\date
\newtoks\Pubnum
\newtoks\pubnum
\Pubnum={~\cr ~\cr YITP/U-\the\pubnum}
\date={\today}
\newcommand{\frontpageskip}{\vspace{12pt plus .5fil minus 2pt}}
\renewcommand{\title}[1]{\frontpageskip
	\begin{center}{\titlefont #1}\end{center}\par}
\renewcommand{\author}[1]{\frontpageskip\par\begin{center}
	{\authorfont #1}\end{center}
	\nobreak
	}
\newcommand{\andauthor}{\frontpageskip\centerline{and}\author}
\newcommand{\authors}{\frontpageskip\noindent}
\newcommand{\address}[1]{\par\begin{center}{\sl #1}\end{center}\par}
\newcommand{\andaddress}{\par\centerline{\sl and}\address}
\renewcommand{\thanks}[1]{\footnote{#1}}
\renewcommand{\abstract}{\par\frontpageskip\centerline{\fourteencp Abstract}
	\vspace{8pt plus 3pt minus 3pt}}
\newcommand\YITP{\address{Uji Research Center,
	       Yukawa Institute for Theoretical Physics\\
               Kyoto University,~Uji 611,~Japan\thanks{Effective
September 20, 1995, Uji Research Center will merge: Yukawa Institute for
Theoretical Physics, Kyoto University, Kyoto 606-01, Japan.}\\}}
\thispagestyle{empty}
%
\pubnum{95-35 \cr UTAP-213/95}
\date{September, 1995}
\titlepage

\baselineskip 0.6cm
\title{\Large\sc Genus Statistics of the Large-Scale Structure\\ with
Non-Gaussian Density Fields}

\author{Takahiko Matsubara}
\address{Department of Physics, The University of Tokyo, Tokyo, 113,
Japan}

\andauthor{ Jun'ichi Yokoyama
}
\YITP

\abstract{
As a statistical measure to quantify the topological structure of the
large-scale structure in the universe, the genus number is calculated
for a number of non-Gaussian distributions in which the density field
is characterized by a nontrivial function of some Gaussian-distributed
random numbers.  As a specific example, the formulae for the lognormal
and the chi-square distributions are derived and compared with the
results of $N$-body simulations together with the previously known
formulae for the Gaussian distribution and second-order perturbation
theory.  It is shown that the lognormal formula fits most
of the simulation data the best.}



\newpage
\newcommand{\gsim}{\mbox{\raisebox{-1.0ex}{$\stackrel{\textstyle >}
{\textstyle \sim}$ }}}
\newcommand{\lsim}{\mbox{\raisebox{-1.0ex}{$\stackrel{\textstyle <}
{\textstyle \sim}$ }}}
\newcommand{\himpc}{{\hbox {$h^{-1}$}{\rm Mpc}} }
\newcommand{\beq}{\begin{equation}}
\newcommand{\eeq}{\end{equation}}
\newcommand{\beqa}{\begin{eqnarray}}
\newcommand{\eeqa}{\end{eqnarray}}
\newcommand{\lmk}{\left(}
\newcommand{\rmk}{\right)}
\newcommand{\lnk}{\left\{ }
\newcommand{\rnk}{\right\} }
\newcommand{\lkk}{\left[}
\newcommand{\rkk}{\right]}
\newcommand{\llangle}{\left\langle}
\newcommand{\rrangle}{\right\rangle}
\newcommand{\bfx}{{\bf x}}
\newcommand{\bfy}{{\bf y}}
\newcommand{\bfz}{{\bf z}}
\newcommand{\rhobar}{\overline{\rho}}
\newcommand{\etal}{ et al.\ }
\newcommand{\dirac}{\delta_{\rm D}}
\newcommand{\etaone}{\eta_1}
\newcommand{\etatwo}{\eta_2}
\newcommand{\etathree}{\eta_3}
\newcommand{\etai}{\eta_i}
\newcommand{\zetaoo}{\zeta_{11}}
\newcommand{\zetatt}{\zeta_{22}}
\newcommand{\zetaot}{\zeta_{12}}
\newcommand{\zetaij}{\zeta_{ij}}
\newcommand{\genus}{G(\nu)}
\newcommand{\sho}{\hat{\sigma}_1}
\newcommand{\sht}{\hat{\sigma}_2}
\newcommand{\shot}{\frac{\hat{\sigma}_1^2}{3}}
\newcommand{\ft}{\tilde{f}}
\newcommand{\at}{\tilde{\alpha}}

\section{Introduction} \label{sec:intro}

One of the most important purposes of observational cosmology is to
probe the power spectrum and the statistical distribution of
primordial fluctuations out of the redshift survey of large-scale
structure, in order to single out the correct model of the evolution
of our universe.  A number of measures have been used to characterize
statistical properties of the large-scale structures.  The most
commonly used quantity is the two-point correlation function
(Totsuji \& Kihara 1969), which is
nothing but the Fourier transform of the power spectrum and does not
contain much information on its statistical distribution.  On the
other hand, the counts-in-cells statistics, including the void
probability (White 1979), directly measure the statistical distribution of
galaxies and various theoretical models based on physical or
mathematical arguments have been proposed to fit the observational
data (Fry 1986).
Although this measure contains mathematically full information
of the statistics in principle, it is difficult to relate the count
analysis with the visual image or connectivity of galaxy clustering
such as filamentary networks, sheet-like or bubble-like structures,
etc..

As a statistical measure to characterize such a topological structure
of galaxy distribution, the genus number has been widely used in the
analysis of recent redshift surveys (Gott, Melott \& Dickinson 1986;
Gott, Weinberg \& Melott 1987; Weinberg, Gott \& Melott 1987; Melott,
Weinberg \& Gott 1988; Gott et al. 1989; Park \& Gott 1991; Park, Gott
\& da Costa 1992; Weinberg \& Cole 1992; Moore et al. 1992; Vogeley et
al. 1994; Rhoads, Gott \& Postman 1994).
Theoretically, however, the value of the genus had been calculated
only  for the random Gaussian field (Adler 1981; Doroshkevich
1970; Bardeen et al.~1986; Hamilton et al.~1986) for a long time
except for the restricted cases such as Rayleigh-L\'evy random-walk fractal
(Hamilton 1988) and union of overlapping balls (Okun 1990). Hence what
one could have done at best was to estimate the genus number on large
scales, which are still in the linear regime, to test the validity of
random Gaussian initial conditions.

The situation was somewhat improved recently because lowest-order
correction to the Gaussian genus number was analytically obtained by
one of us using the multi-dimensional Edgeworth expansion around the
Gaussian distribution (Matsubara 1994).  Detailed comparison has also
been done with the results of $N$-body simulations and it has been
shown that the new formula fits the numerical data well in the
semi-linear regime but not in the nonlinear regime (Matsubara \& Suto
1995).  This is in accord with the fact that the one-point probability
distribution function (PDF) based on the Edgeworth series ceases to
fit the counts-in-cells once the root-mean-square (rms) value of the
density contrast becomes as large as $\simeq 1/4$ (Juszkiewicz et
al. 1995; Ueda \& Yokoyama 1995).  Thus it is also desirable to find
an analytic expression of the genus number for realistic non-Gaussian
distributions just as various non-Gaussian models have been proposed
for counts-in-cells statistics (Fry 1986).

In the present paper we present analytic formulae of genus number for
some non-Gaussian distributions in which the density contrast is given
by a nontrivial function of either a single Gaussian-distributed
random number or its combinations.  The former includes the lognormal
distribution which has been used as a model of galaxy distribution
ever since Hubble (1934). The lognormal model is found to fit the
three-dimensional PDF of observed galaxies (Hamilton 1985; Bouchet et
al.~1993; Kofman et al.~1994), and of cell-counts in CDM-type $N$-body
simulations (Kofman et al.~1994; Ueda \& Yokoyama 1995). Coles \&
Jones (1991) argued for the lognormal mapping of the linear density
field to describe its nonlinear evolution in the universe. The latter
includes chi-square distribution which is closely related with the
negative binomial distribution, another widely used distribution with
a hierarchical property of higher-order cumulants (Fry 1986;
Carruthers 1991; Gazta\~naga \& Yokoyama 1993; Bouchet et al. 1993).
We then compare the new formulae with $N$-body simulation data with
various initial spectra as well as the previously known
formulae for the Gaussian distribution and second-order perturbation
theory.

The rest of the paper is organized as follows.  In \S
\ref{sec:gauss} we review derivation of the genus number in Gaussian
distribution and its lowest-order correction arising from the
three-point correlation function.  In \S \ref{sec:log} the genus
number is given in the case the density contrast is given by a
nontrivial but monotonic function of a Gaussian-distributed quantity.
Then in \S \ref{sec:mult} it is extended to the case the density
fluctuation is a function of several independent Gaussian variables.
In \S \ref{sec:nbody} various analytic formulae of the genus and the
PDF are compared with numerical results obtained from $N$-body simulations.
Finally \S \ref{sec:concl} is devoted to discussion and conclusions.

\section{Genus Curve in Gaussian Distribution and its Nonlinear Correction
through the Edgeworth Series} \label{sec:gauss}

The genus curve $\genus$ is defined by $-1/2$ times the total Euler
number per unit volume of isodensity contours of a continuous density
field.  It corresponds to
\[
\rm \lkk \# (holes)- \# (isolated~ regions) \rkk/ volume.
\]
Here the density threshold of the contour is specified by $\nu \equiv
\delta/\sqrt{\langle\delta^2\rangle}$ with
$\delta$ being the density contrast which is smoothed appropriately.
Mathematically, it is given by the following expectation value
(Doroshkevich 1970; Adler 1981; Bardeen et al. 1986).
\beq
\genus=-\frac{1}{2}
\llangle \dirac\lmk \delta(\bfx)-\nu\sigma\rmk\dirac(\etaone)
\dirac(\etatwo)|\etathree|(\zetaoo\zetatt-\zetaot^2)\rrangle,
\label{genusdef}
\eeq
where $\etai\equiv\partial_i\delta(\bfx)\equiv \delta_{,i}(\bfx)$,
$\zetaij\equiv\partial_i\partial_j\delta(\bfx)\equiv
\delta_{,ij}(\bfx)$ and $\dirac(\cdot)$ is Dirac's delta function.
Thus it can in principle be calculated once a seven-point probability
distribution function of $\delta(\bfx)$ is known.

In the case $\delta(\bfx)$ obeys Gaussian distribution, we find
$\delta(\bfx),~\eta_i(\bfx),$ and $\zeta_{jk}(\bfx)$ are also
Gaussian-distributed with a vanishing mean and two-body correlations
given by
\beqa
 \langle\delta^2(\bfx)\rangle &\equiv& \sigma^2,
 ~~~\langle\delta(\bfx)\eta_i(\bfx)\rangle=0,~~~
 \langle\delta(\bfx)\zeta_{ij}(\bfx)\rangle=-\frac{\sigma_1^2}{3}\delta_{ij}
 \nonumber\\
 \langle\eta_i(\bfx)\eta_j(\bfx)\rangle&=&\frac{\sigma_1^2}{3}\delta_{ij},~~~
 \langle\eta_i(\bfx)\zeta_{jk}(\bfx)\rangle=0,\\
 \langle\zeta_{ij}(\bfx)\zeta_{kl}(\bfx)\rangle&=&
 \frac{\sigma_2^2}{15}(\delta_{ij}\delta_{kl}+\delta_{ik}\delta_{jl}
 +\delta_{il}\delta_{jk}),\nonumber
\eeqa
where $\sigma^2$, $\sigma_1^2$ and $\sigma_2^2$ are defined,
respectively, by
\beq
  \label{eq:sigma}
  \sigma^2\equiv \langle\delta^2\rangle,~~~
  \sigma_1^2\equiv \langle(\nabla\delta)^2\rangle,~~~{\rm and}~~~
  \sigma_2^2\equiv \langle(\nabla^2\delta)^2\rangle.
\eeq
The final result is
\beq
 \label{eq:lin}
 \genus=\frac{e^{-\frac{\nu^2}{2}}}{(2\pi)^2}
\lmk\frac{\sigma^2_1}{3\sigma^2}\rmk^{3/2}
 (1-\nu^2)\equiv G_{\rm RG}(\nu).
\eeq

If the initial fluctuation is a Gaussian random field, linear theory
predicts that the genus is described by this Gaussian formula
(\ref{eq:lin}) and the PDF by
\begin{equation}
   P_{\rm RG}(\nu) = \frac{e^{-\nu^2/2}}{\sqrt{2\pi}}.
   \label{eq:pdflin}
\end{equation}

Next we consider correction due to the presence of higher-order
correlation functions, which arises as a result of nonlinear
gravitational evolution, in terms of the multi-dimensional Edgeworth
expansion following Matsubara (1994) and Matsubara \& Suto (1995). The
final result is
\begin{equation}
   G_{\rm 2nd}(\nu) = - \frac{e^{-\frac{\nu^2}{2}}}{(2\pi)^2}
   \left(\frac{\sigma_1^2}{3\sigma^2}\right)^{\frac{3}{2}}
   \left[ H_2(\nu)
          + \sigma \left( \frac{S}{6} H_5(\nu)
                          + \frac{3T}{2} H_3(\nu)
                          + 3U H_1(\nu)\right)
          + {\cal O}(\sigma^2)
   \right] .
   \label{eq:second}
\end{equation}
In the above expression, $H_n(\nu) \equiv (-)^n e^{\nu^2/2} (d/d\nu)^n
e^{-\nu^2/2}$ is the $n$-th order Hermite polynomial, and $S$, $T$,
and $U$ are defined as
\begin{eqnarray}
   && S = \frac{1}{\sigma^4} \langle\delta^3\rangle,
   \nonumber \\
   && T = - \frac{1}{2\sigma_1^2\sigma^2}
            \langle \delta^2 \nabla^2 \delta \rangle,
   \label{eq6} \\
   && U = - \frac{3}{4\sigma_1^4}
            \langle \nabla\delta\cdot\nabla\delta
                    \nabla^2 \delta \rangle ,
   \nonumber
\end{eqnarray}
respectively, which we call generalized skewness. They come from the
three-point correlation function that can be evaluated also by
second-order perturbation theory (Matsubara 1994). One should note
that generalized skewness should be evaluated for smoothed
fluctuations that are evolved nonlinearly. Only after taking into
account such smoothing effect, one can compare the second-order
perturbation theory with observations.  If one uses the Gaussian
window with the smoothing length $R$, and assumes the Gaussian initial
fluctuations, they are explicitly computed as
\begin{eqnarray}
   && S = \frac{1}{4\pi^4}
   \left[ (2 + K) L_{220} + 3 L_{131} + (1 - K) L_{222}\right],
\nonumber \\
   && T = \frac{1}{60\pi^4}
   [5(5 + 2K) L_{240} + 3(9 + K) L_{331} + 15 L_{151}
\nonumber \\
   && \hskip5.0cm + 10(2 - K) L_{242} + 3(1 - K) L_{333}],
   \label{eq8} \\
   && U = \frac{1}{140\pi^4}
   \left[7(3 + 2K) L_{440} + 21 L_{351} - 5(3 + 4K) L_{442} -
   21 L_{353} - 6(1 - K) L_{444}\right].
\nonumber
\end{eqnarray}
Here $L_{\alpha\beta n}(R)$ stands for the following integral
\begin{eqnarray}
   &&
   L_{\alpha\beta n}(R) \equiv
   \frac{\sigma_1^{4 - \alpha - \beta}}{\sigma^{8 - \alpha - \beta}}
   \int_0^\infty dx \int_0^\infty dy \int_{-1}^1 d\mu
   e^{- R^2(x^2 + y^2 + \mu x y)} x^\alpha y^\beta P_n(\mu)
   P(x) P(y)
   \label{eq11}
   \\
   && 
   = (-)^n \sqrt{2\pi}
   \frac{\sigma_1^{4 - \alpha - \beta}}{R\sigma^{8 - \alpha - \beta}}
   \int_0^\infty dx \int_0^\infty dy
   e^{- R^2(x^2 + y^2)} x^{\alpha-1/2} y^{\beta-1/2}
   I_{n+1/2}(xyR^2) P(x) P(y),
   \nonumber\\
   \label{eq12}
\end{eqnarray}
where $\sigma$ and $\sigma_1$ are defined by (\ref{eq:sigma}) in which
$\delta$ is the Gaussian smoothed density fluctuation of linear theory
over the scale $R$, $P_n$ is the $n$-th order Legendre polynomial, and
$I_\nu$ is a modified Bessel function.  The above results (\ref{eq8})
to (\ref{eq12}) hold for arbitrary values of density parameter
$\Omega$ and cosmological constant $\Lambda$. The latter effect
manifests only through the function $K = K(\Omega,\lambda)$ which very
weakly depends on $\Omega$ and $\lambda$ (Bouchet et al.~1992;
Bernardeau 1994) where $\lambda \equiv \Lambda/(3H^2)$, $H$ is the
Hubble parameter.  Its explicit form for $K$ has been derived by
Matsubara (1995) as
\begin{eqnarray}
   K(\Omega,\lambda) =
   \frac{\Omega}{4} - \frac{\lambda}{2} -
   \left(\int_0^1 dx X^{-3/2}\right)^{-1} +
   \frac32 \left(\int_0^1 dx X^{-3/2}\right)^{-2}
   \int_0^1 dx X^{-5/2} ,
   \label{eq9}
\end{eqnarray}
where
\begin{eqnarray}
   X(x) \equiv \Omega/x + \lambda x^2 + 1 - \Omega - \lambda .
   \label{eq10}
\end{eqnarray}
In the two specific models we adopt below, we find
$K(1,0) = 3/7 = 0.4286$ and $K(0.2,0.8) = 0.4335$.

For the power-law fluctuation spectra $P(k) \propto k^n$, $S$, $T$
and $U$ can be written down explicitly in terms of the hypergeometric
function as
\begin{eqnarray}
   &&
   S = 3 F\left(\frac{n+3}{2},\frac{n+3}{2},\frac{3}{2};\frac{1}{4}\right)
   - (n + 2 - 2K)
   F\left(\frac{n+3}{2},\frac{n+3}{2}, \frac{5}{2};\frac{1}{4}\right),
   \nonumber \\
   &&
   T = 3F\left(\frac{n+3}{2},\frac{n+5}{2}, \frac{3}{2};\frac{1}{4}\right)
   - (n + 3 - K)
   F\left(\frac{n+3}{2},\frac{n+5}{2},\frac{5}{2};\frac{1}{4}\right)
   \nonumber \\
   && \hskip4.0cm
   + \frac{(n-2)(1-K)}{15}
   F\left(\frac{n+3}{2},\frac{n+5}{2},\frac{7}{2};\frac{1}{4}\right) ,
   \label{eq:powerstu} \\
   &&
   U = F\left(\frac{n+5}{2},\frac{n+5}{2},\frac{5}{2};\frac{1}{4}\right)
   - \frac{n + 4 - 4K}{5}
   F\left(\frac{n+5}{2},\frac{n+5}{2},\frac{7}{2};\frac{1}{4}\right).
   \nonumber
\end{eqnarray}
The expressions for $S$ in equations (\ref{eq8}) and (\ref{eq:powerstu}) are
derived by {\L}okas et al.~(1994) which are equivalent to the other
form independently derived by Matsubara (1994). Similarly we transform
the expressions for $T$ and $U$ presented in Matsubara (1994;
eqs.~[16] and [18]) using the function $L_{\alpha\beta n}(R)$, which
are given in equations (\ref{eq8}) and (\ref{eq:powerstu}).

This result of equation (\ref{eq:second}) is the analog of the
second-order Edgeworth series of PDF (Juszkiewicz et al. 1995;
Bernardeau \& Kofman 1995):
\begin{equation}
   P_{\rm 2nd}(\nu) = \frac{e^{-\nu^2/2}}{\sqrt{2\pi}}
   \left[1 + \sigma \frac{S}{6} H_3(\nu) + {\cal O}(\sigma^2) \right]
   \label{eq:pdfsecond}
\end{equation}
\vskip 4cm
\section{The Case Density Field is a Monotonic Function of a
Random Gaussian Variable} \label{sec:log}
Next we turn to evaluation of $\genus$ for a non-Gaussian distribution
whose statistical properties are characterized by a monotonic function
$F$ as
\beq
   \delta(\bfx)=F[\phi(\bfx)],
\eeq
with $\phi(\bfx)$ being a Gaussian-distributed random field with
vanishing mean and unit variance, so that
one-point PDF of $\delta$ reads
\beq
  P[\delta]d\delta= \frac{1}{\sqrt{2\pi}|F'[F^{-1}(\delta)]|}
  \exp\lnk -\frac{1}{2}\lkk F^{-1}(\delta)\rkk^2\rnk d\delta.
\eeq
In this case we find
\beqa
  \etai(\bfx) &=& F'[\phi(\bfx)]\phi_{,i}(\bfx),\\
  \zetaij(\bfx)&=&
  F''[\phi(\bfx)]\phi_{,i}(\bfx)\phi_{,j}(\bfx)
  +F'[\phi(\bfx)]\phi_{,ij}(\bfx).\nonumber
\eeqa
Therefore $\genus$ is calculated as
\beqa
 && \genus = -\frac{1}{2}\llangle\dirac\lmk F(\phi)-\nu\sigma\rmk
  \dirac\lmk F'(\phi)\phi_{,1}\rmk\rule{0mm}{9mm}
  \dirac\lmk F'(\phi)\phi_{,2}\rmk
  |F'(\phi)\phi_{,3}|\right. \nonumber\\*
 &&\!\!\!\!\times\left.\lnk\lkk F''(\phi)\phi_{,1}^2+F'(\phi)\phi_{,11}\rkk
      \lkk F''(\phi)\phi_{,2}^2+F'(\phi)\phi_{,22}\rkk
     - \lkk F''(\phi)\phi_{,1}\phi_{,2}
    +F'(\phi)\phi_{,12}\rkk^2\rnk\rule{0mm}{9mm}\rrangle_\phi
     \nonumber\\*
  &=&
  -\frac{1}{2}\llangle \frac{1}{|F'(\phi)|}
  \dirac\lmk \phi-F^{-1}(\nu\sigma)\rmk
  \frac{\dirac(\phi_{,1})}{|F'(\phi)|}
  \frac{\dirac(\phi_{,2})}{|F'(\phi)|}
  |F'(\phi)||\phi_{,3}|
  F'(\phi)^2(\phi_{,11}\phi_{,22}-\phi_{,12}^2)\rrangle_\phi
  \nonumber\\*
  &=& \frac{1}{(2\pi)^2}
  \lmk\frac{\langle(\nabla\phi)^2\rangle_\phi}{3}\rmk^{3/2}
  \exp\lnk-\frac{\lkk F^{-1}(\nu\sigma)\rkk^2}{2}\rnk
  \lnk 1-\lkk F^{-1}(\nu\sigma)\rkk^2\rnk,
\eeqa
where we have made use of the assumption that $F(\phi)$ is monotonic with
nonvanishing $F'(\phi)$.

Thus the shape of the genus curve is obtained from that of the
Gaussian distribution simply replacing $\nu$ by $F^{-1}(\nu\sigma)$.
This is as expected because mapping in terms of a monotonic function
does not change the shape of the density profile.  Using the relations
$\sigma^2=\langle F(\phi)^2\rangle_\phi$ and $\sigma_1^2=\langle
F'(\phi)^2\rangle_\phi\langle (\nabla\phi)^2\rangle_\phi$ for
isotropic spaces, the overall factor can be rewritten to yield the
final result:
\beq
  \genus=\lmk\frac{\langle F(\phi)^2\rangle_\phi}{\langle
  F'(\phi)^2\rangle_\phi}\rmk^{3/2} G_{\rm RG}\lmk
  F^{-1}(\nu\sigma)\rmk.
\eeq

In the particular case of the lognormal distribution, the function $F$
is defined by
\beq
  F[\phi(\bfx)]=\frac{1}{\sqrt{1+\sigma^2}}\exp\lkk \sqrt{\ln(1+\sigma^2)}
  \phi(\bfx)\rkk -1.
\eeq
We therefore find the PDF in the lognormal distribution as
\begin{equation}
   P_{\rm LN}(\nu) =
   \frac{\sigma}{(1 + \nu\sigma)\sqrt{2\pi \ln(1 + \sigma^2)}}
   \exp\left(
      - \frac{\left\{\ln\left[
         (1 + \nu\sigma)\sqrt{1 + \sigma^2}\right]\right\}^2}
        {2 \ln(1 + \sigma^2)}
      \right),
   \label{eq:pdfln}
\end{equation}
and the genus curve in the lognormal distribution as
\beqa
  \label{eq:logn}
  G_{\rm LN}(\nu)&=&\frac{1}{(2\pi)^2}
  \frac{\sigma_1^3}{\lkk 3(1+\sigma^2)\ln(1+\sigma^2)\rkk^{3/2}}
  \nonumber\\ &\times&
  \exp\lmk -\frac{\lnk\ln\lkk(1+\nu\sigma)\sqrt{1+\sigma^2}\rkk\rnk^2}
  {2\ln(1+\sigma^2)}\rmk \lmk 1-
  \frac{\lnk\ln\lkk(1+\nu\sigma)\sqrt{1+\sigma^2}\rkk\rnk^2}
  {\ln(1+\sigma^2)} \rmk.
\eeqa
One can easily check that the above expression reduces to the Gaussian
formula in the limit $\sigma \longrightarrow 0$.

\section{The Case Density Fields Depends on Multiple Gaussian Fields}
\label{sec:mult}

\subsection{General consideration}

We now extend the above analysis to the case statistical properties
of the density field is characterized by a number of independent
Gaussian fields $\alpha^a(\bfx)$ through a function $f$ as
\beq
  \delta(\bfx)=f[\alpha^1(\bfx),\alpha^2(\bfx),...,\alpha^n(\bfx)].
\eeq
Here we assume that $\alpha^a$'s are mutually independent and all have
vanishing mean and unit variance.  Then introducing new variables
$\beta_i^a\equiv\alpha_{,i}^a$ and $\gamma_{ij}^a\equiv
\alpha_{,ij}^a$ we find
\beqa
   \eta_i&=&\frac{\partial f}{\partial \alpha^a}\alpha^a_{,i}
   \equiv f_a\beta^a_i,  \nonumber\\
   \zeta_{ij}&=&\frac{\partial^2 f}{\partial\alpha^a\partial\alpha^b}
   \beta^a_i\beta^b_j + \frac{\partial f}{\partial \alpha^a_i}
   \alpha^a_{,ij} \equiv f_{ab}\beta^a_i\beta^b_j
   + f_a\gamma^b_{ij}.
\eeqa
Here and below summation over repeated Latin indices {\it a,b,c,...,h}
is implicitly assumed.
Then the genus number is formally written as
\beqa
\genus&=&-\frac{1}{2}\llangle\dirac(f-\nu\sigma)\dirac(f_e\beta^e_1)
\dirac(f_g\beta^g_2)|f_h\beta_3^h|\right.  \nonumber\\
&~&~~~\times\left.\lkk(f_{ab}\beta^a_1\beta^b_1+f_a\gamma^a_{11})
(f_{cd}\beta^c_2\beta^d_2+f_c\gamma^c_{22})-
(f_{ab}\beta^a_1\beta^b_2+f_a\gamma^a_{12})^2\rkk\rrangle.
\eeqa
Here the two-body correlations are given by
\beqa
 \langle\alpha^a\alpha^b\rangle&=&\delta^{ab},~~~
 \langle\alpha^a\beta_i^b\rangle=0,~~~
 \langle\alpha^a\gamma_{ij}^a\rangle=-\frac{\sho^2}{3}\delta^{ab}\delta_{ij},
 \nonumber\\
 \langle\beta_i^a\beta^b_j\rangle&=&\frac{\sho^2}{3}\delta^{ab}\delta_{ij},
 ~~~\langle\beta^a_i\gamma^b_{jk}\rangle=0,\nonumber\\
 \langle\gamma^a_{ij}\gamma^b_{kl}\rangle
 &=&\frac{\sht^2}{15}\delta^{ab}(\delta_{ij}\delta_{kl}
 +\delta_{ik}\delta_{jl}+\delta_{il}\delta_{jk}),
\eeqa
where $\sho$ and $\sht$ are defined by
\beq
  \sho^2 \equiv  \langle(\nabla\alpha^p)^2\rangle,~~~{\rm and}~~~
  \sht^2 \equiv  \langle(\nabla^2\alpha^p)^2\rangle.
\eeq

Introducing new variables $\omega^a_{ij}\equiv
\gamma^a_{ij}+\sho^2\delta_{ij}\alpha^a/3$ we find $\alpha^a$, $\beta^b_i$,
and $\omega^c_{jk}$ are random-Gaussian variables whose correlations are
totally decoupled from each other:
\beqa
 \langle\alpha^a\omega^b_{jk}\rangle&=&\langle\beta^a_i\omega^b_{jk}\rangle
 =0, \nonumber\\
 \langle\omega^a_{ij}\omega^b_{kl}\rangle&=&\delta^{ab}
 \lkk\lmk \frac{\sht^2}{15}-\frac{\sho^4}{9}\rmk\delta_{ij}\delta_{kl}
 +\frac{\sht^2}{15}(\delta_{ik}\delta_{jl}+\delta_{il}\delta_{jk})\rkk.
\label{omegacorrelation}
\eeqa
The above fact greatly simplifies the subsequent averaging procedures,
because we can average over $\omega^a_{ij}$, $\beta_3^b$, $\beta_1^c$,
and $\beta_2^d$ in turn independently as follows.

First, in
\beqa
&&\!\!\!\!\genus=-\frac{1}{2}\llangle\rule{0mm}{9mm}
\dirac(f-\nu\sigma)\dirac(f_e\beta^e_1)
\dirac(f_g\beta^g_2)|f_h\beta_3^h|\right.  \\
&&\!\!\!\!\!\!\!\!\times\lnk\lkk f_a\lmk\omega^a_{11}-\shot\alpha^a\rmk
 +f_{ab}\beta_1^a\beta_1^b\rkk
 \lkk f_c\lmk\omega^c_{22}-\shot\alpha^c\rmk
 +f_{cd}\beta_2^c\beta_2^d\rkk
-\left.\lkk\rule{0mm}{5mm} f_a\omega^a_{12}+f_{ab}\beta_1^a\beta_2^b\rkk^2\rnk
 \rrangle, \nonumber
\eeqa
averaging over $\omega_{ij}^a$ can be readily done  using
(\ref{omegacorrelation}).  The expression in the curly bracket
should be replaced by
\beq
\llangle\lnk\rule{0mm}{4mm}\cdots\rnk\rrangle_\omega
=-\frac{\sho^4}{9}f_af_b(\delta^{ab}-\alpha^a\alpha^b)
 -\shot f_{ab}(\beta^a_1\beta^b_1+\beta^a_2\beta^b_2)f_c\alpha^c
 +f_{ab}f_{cd}(\beta^a_1\beta^b_1\beta^c_2\beta^d_2-
   \beta^a_1\beta^b_2\beta^c_1\beta^d_2).
\eeq
Next $\beta_3^a-$average is performed noting that the linear
combination $f_h\beta_3^h$ is also a Gaussian with a vanishing mean and
the variance
\beq
  \langle (f_h\beta_3^h)^2\rangle_{\beta_3}=\shot\ft^2,~~~~~
  {\rm with}~~\ft^2\equiv\delta^{ab}f_a f_b,
\eeq
to yield
\beq
  \langle|f_h\beta_3^h|\rangle_{\beta_3}=\sqrt{\frac{2}{3\pi}}\ft\sho.
\eeq
We assume only nonvanishing $\ft$ contributes to the final result.
The averaging over $\beta_1^a$ or $\beta_2^a$ is more involved but it
can be done utilizing the fact that
$u_i\equiv f_a \beta_i^a,~ \beta_i^p,$ and $\beta_i^q$ $(p\neq q)$
constitute a trivariate Gaussian distribution with the correlation matrix
\beqa
  M= \shot\lkk \begin{array}{ccc}
            \ft^2 & f_p & f_q \\
            f_p   & 1   & 0    \\
            f_q   & 0   & 1    \end{array} \rkk.
\eeqa
After some straight forward calculations we find
\beqa
 \langle \dirac(u_i)\rangle_{\beta_i}
 &=&\sqrt{\frac{3}{2\pi}}\frac{1}{\ft\sho}, \\
 \langle \dirac(u_i)\beta^p_i\beta^q_i\rangle_{\beta_i}
 &=&\sqrt{\frac{3}{2\pi}}\frac{1}{\ft\sho}\times
  \frac{\ft^2\delta_{pq}-f_pf_q}{3\ft^2}\sho^2,
\eeqa
including the case with $p=q$.
We thus obtain the expression for $\genus$ leaving only
$\alpha-$average as
\beqa
\genus&=&-\frac{\sho^3}{(6\pi)^{3/2}}
\llangle\dirac(f-\nu\sigma)\frac{1}{\ft}\lkk (f_a\alpha^a)^2-\ft^2
-2\lmk f_{aa}-\frac{f_{ab}f_a f_b}{\ft^2}\rmk f_c\alpha^c
+(f_{aa})^2
\right.\right. \nonumber\\ &~&~~~~~~~~~~~~~~~~~ \left.\left.
-f_{ab}f_{ab}-
\frac{2}{\ft^2}(f_{ab}f_af_bf_{cc}-f_{ac}f_{bc}f_af_b)\rkk
\rrangle_{\alpha}.  \label{alphaav}
\eeqa
This average cannot be calculated until we specify the function
$f(\alpha^a)$.  We therefore move on to a specific example of the
chi-square distribution in the next subsection.

\subsection{Chi-square distribution}

In  chi-square matter distribution the density field, $\rho(\bfx)$, is
given by
\beq
  \rho(\bfx)=\frac{\rhobar}{n}\sum_{p=1}^{n}
  \lkk\alpha^p(\bfx)\rkk^2,
\eeq
with $\rhobar$ being the mean density.  The density contrast
reads
\beq
  \delta(\bfx)=\frac{\rho(\bfx)-\rhobar}{\rhobar}
  = \frac{1}{n}\sum_{p=1}^{n}\lkk\alpha^p(\bfx)\rkk^2 -1.
\eeq
Hence we find
\beq
  \sigma^2=\langle\delta(\bfx)^2\rangle=\frac{2}{n}. \label{nsigma}
\eeq
Although $n$ is a positive integer by definition, we can perform
analytic continuation to arbitrary positive number and replace $n$ by
$2/\sigma^2$ using (\ref{nsigma}) in what follows.

Thus in the chi-square distribution the function $f(\alpha)$ is given
by
\beq
  f(\alpha)=\frac{\sigma^2}{2}\alpha^a\alpha^a-1
   \equiv \frac{\sigma^2}{2}\at^2-1,
\eeq
with $f_a=\sigma^2\alpha^a$ and $f_{ab}=\sigma^2\delta^{ab}$.
We therefore find from (\ref{alphaav}),
\beqa
  \genus&=&-\frac{\sho^3}{(6\pi)^{3/2}}
\llangle\dirac\lmk f(\at)-\nu\sigma\rmk\frac{1}{\sigma^2\at}
\lkk\sigma^4\at^4+\sigma^4\at^2-4\sigma^2\at^2+2\sigma^4-6\sigma^2+4\rkk
\rrangle_\alpha \nonumber\\
&=& \frac{\sho^3}{(3\pi)^{3/2}}\frac{1}{\sqrt{1+\nu\sigma}}
\lmk 1-\nu^2-\frac{\sigma(\nu+\sigma)}{2}\rmk P_{\rm CH}(\nu)
\equiv G_{\rm CH}(\nu).
\eeqa
Here $P_{\rm CH}(\nu)$ is the one-point PDF of $\delta(\bfx)/\sigma$
as a function of the threshold $\nu$, which is calculated as
\beqa
 P_{\rm CH}(\nu)&\equiv&\langle\dirac\lmk f(\at)/\sigma-\nu\rmk\rangle
 \nonumber\\ &=&\int\frac{d^n\alpha^a}{(2\pi)^{n/2}}e^{-\at^2/2}
 \dirac\lmk \frac{\sigma}{2}\at^2-\frac{1}{\sigma}-\nu\rmk \nonumber\\
 &=& \frac{(1+\nu\sigma)^{n/2-1}}{\sigma^{n-1}\Gamma(n/2)}
 \exp\lmk-\frac{1+\nu\sigma}{\sigma^2}\rmk \label{onepoint}\\ &=&
 \frac{(1+\nu\sigma)^{\sigma^{-2}-1}}{\sigma^{2\sigma^{-2}-1}
 \Gamma(\sigma^{-2})}
 \exp\lmk-\frac{1+\nu\sigma}{\sigma^2}\rmk. \nonumber
\eeqa
Using the relation
\beq
 \sigma_1^2=3\langle\eta_1^2\rangle=3\langle f_a\beta_1^a\rangle
 =\sigma^4\sho^2 n=2\sigma^2\sho^2,
\eeq
we finally obtain
\beq
  G_{\rm CH}(\nu)=\frac{1}{(2\pi)^{3/2}}\lmk\frac{\sigma_1^2}
  {3\sigma^2}\rmk^{3/2}
  \lmk 1-\nu^2-\frac{\sigma(\nu+\sigma)}{2}\rmk
  \frac{(1+\nu\sigma)^{\sigma^{-2}-3/2}}
  {\Gamma(\sigma^{-2})\sigma^{2\sigma^{-2}-1}}
  \exp\lmk-\frac{1+\nu\sigma}{\sigma^2}\rmk.  \label{chisquare}
\eeq
Since (\ref{onepoint}) reduces to the Gaussian distribution
for $\sigma\longrightarrow 0$ (\ref{chisquare}) also
coincides with  (\ref{eq:lin}) in this limit.

Note that (\ref{onepoint}) indicates that one-point PDF of the density
field obeys the negative binomial distribution
\beq
 P_{\rm NB}[\rho(\bfx)]d\rho=
 \frac{1}{\Gamma(\sigma^{-2})\sigma^2}\lmk\frac{\rho}{\rhobar\sigma^2}
 \rmk^{\sigma^{-2}-1}\!\!\!\!\!\!\!\!\exp\lmk-\frac{\rho}{\rhobar\sigma^2}\rmk
 \frac{d\rho}{\rhobar}.
\eeq
The negative binomial distribution is often used to fit the data of
counts-in-cells analysis, namely, as a model of a PDF of smoothed
density field.  A number of observational analyses have shown that it
fits the counts-in-cells histogram well for some volume-limited
redshift samples (Gazta\~naga \& Yokoyama 1993; Bouchet et al.\ 1993).
  We also note that this distribution has a
hierarchical property of higher-order connected moments.  That is,
$N-$th order cumulant $\kappa_N$ is given by
$\kappa_N=(N-1)!(\kappa_2)^{N-1}= (N-1)!\sigma^{2N-2}$.

On the other hand, in spite of the above similarity, the chi-square
distribution and the negative binomial distribution are essentially
different from each other once spatial dependence is taken into account.
To illustrate it let us consider two- and three-point correlation
functions, $\xi$ and $\zeta$, respectively, in the chi-square distribution.
The former is calculated as
\beq
  \xi(\bfx,\bfy)\equiv\langle\delta(\bfx)\delta(\bfy)\rangle
  =\frac{\sigma^4}{4}\langle\alpha^a(\bfx)\alpha^a(\bfx)
   \alpha^b(\bfy)\alpha^b(\bfy)\rangle -1
  =\sigma^2 w^2(\bfx,\bfy),
\eeq
where $\langle\alpha^a(\bfx)\alpha^b(\bfy)\rangle \equiv \delta^{ab}
w(\bfx,\bfy)$. Similarly, the latter is given by
\beqa
 \zeta(\bfx,\bfy,\bfz)&\equiv&\langle\delta(\bfx)\delta(\bfy)\delta(\bfz)
 \rangle \nonumber\\
 &=&2\sigma^4 w(\bfx,\bfy) w(\bfy,\bfz) w(\bfz,\bfx)\\
 &=&2\sigma\lkk\xi(\bfx,\bfy)\xi(\bfy,\bfz)\xi(\bfz,\bfx)
  \rkk^{1/2}, \nonumber
\eeqa
which is very different from a hierarchical form.
On the other hand,
if the negative binomial distribution would fit the counts-in-cells
analysis for any shape and size of the sampling cell, we would expect
the following relation to hold
\beq
  \zeta(\bfx,\bfy,\bfz) \cong
  \frac{2}{3}\lkk\xi(\bfy,\bfz)\xi(\bfz,\bfx)+\xi(\bfz,\bfx)\xi(\bfx,\bfy)+
   \xi(\bfx,\bfy)\xi(\bfy,\bfz)\rkk,
\eeq
with the observed slope (Peebles \& Groth 1975; Groth \& Peebles 1977)
of the power-law two-point
correlation function (Gazta\~naga \& Yokoyama 1993).
Thus the two distribution should
be distinguished from each other even if one-point PDF has exactly the
same form.

\section{Comparison with Numerical Simulations} \label{sec:nbody}

We measure the genus of the four data sets from cosmological $N$-body
simulations with random-Gaussian initial conditions, kindly provided
by T.~Suginohara and Y.~Suto. Three models are evolved in the
Einstein-de Sitter universe with the scale-free initial fluctuation
spectra (at expansion factor $a=1.0$):
\begin{equation}
   P(k) \propto k^n \qquad (n= -1, \, 0, \, {\rm and} \, 1) .
\end{equation}
The last model corresponds to a spatially-flat low-density cold dark
matter (LCDM) model.  In this specific example, we assume $\Omega_0 =
0.2$, $\lambda_0 = 0.8$, and $h = 1.0$ (Suginohara \& Suto 1991).  The
amplitude of the power spectrum in the LCDM model at $a=6$ is
normalized so that the top-hat smoothed rms mass fluctuation is unity
at $8 \himpc$.  In fact this LCDM model can be regarded to represent a
specific example of the most successful cosmological scenarios so far
(e.g., Suto 1993).  All models are evolved with a hierarchical tree
code implementing the fully periodic boundary condition in a cubic
volume of $L^3$. The physical comoving size of the computational box
in the LCDM model is $L=100 h^{-1}$Mpc.  The number of particles
employed in the simulations is $N = 64^3$, and the gravitational
softening length is $\epsilon_g = L/1280$ in comoving.  Further
details of the simulation models and other extensive analyses are
described in Hernquist, Bouchet \& Suto (1991), Suginohara et
al. (1991), Suginohara \& Suto (1991), Suto (1993), Matsubara \& Suto
(1994), and Suto \& Matsubara (1994).

The computation of the genus from the particle data is performed using
the code kindly provided by David Weinberg (Weinberg 1988; Gott et al.
1989). In short the procedure goes as follows; (i) the computational
box is divided into $N_c^3 (=128^3)$ cubes, and the density
$\rho_g({\bf r})$ at the center of each cell is computed using
Cloud-In-Cell density assignment. (ii) the Fourier-transform:
\begin{equation}
 \tilde{\rho}_g({\bf k}) \equiv {1 \over L^3}
\int \rho_g({\bf r}) {\rm exp} (  i{\bf k} \cdot{\bf r})  d^3r ,
\end{equation}
is convolved with the Gaussian filter, and transformed back to define
a {\it smoothed} density of each cell (with the filtering length
$R_f$):
\begin{equation}
\rho_s({\bf r}; R_f) = {L^3 \over 8\pi^3} \int \tilde{\rho}_g({\bf k})
\exp \lmk - \frac{k^2 R_f^2}{2} - i{\bf k} \cdot{\bf r}\rmk  d^3k .
\end{equation}
(iii) the rms amplitude of the density fluctuations is computed
directly from the smoothed density:
\begin{equation}
\sigma(R_f) \equiv \sqrt{\langle
                     (\rho_s/\bar\rho -1)^2 \rangle} ,
\end{equation}
where $\bar\rho$ is the mean density of the particles. (iv) The
isodensity surface of the critical density:
\begin{equation}
\rho_c \equiv \left[ 1 + \nu \sigma(R_f) \right] \bar\rho
\label{eq:rhoc}
\end{equation}
is approximated by the boundary surface of the high-density
($\rho_s>\rho_c$) and low-density ($\rho_s<\rho_c$) cells.  (v) Then
the genus of the surface is computed by summing up the angle deficit
$D(i,j,k)$ at the vertex of cell $(i,j,k)$:
\begin{equation}
g_s(\nu) = - { 1\over 4\pi} \sum_{i,j,k =1}^{N_c} D(i,j,k)  .
\end{equation}
The way to compute $D(i,j,k)$ is detailed in Gott et al. (1986).  The
genus curve $G(\nu)$ is defined to be the number of genus per
unit volume as a function of the threshold $\nu$.  (vi) We repeated
the above procedure 50 times using the bootstrap resampling method
(Ling, Frenk \& Barrow 1986) in order to estimate the statistical
errors of $G(\nu)$.

It should be noted that earlier papers (e.g., Gott et al. 1989;
Rhoads, Gott, \& Postman 1994; Vogeley et al. 1994) {\it defined} the
density threshold $\nu$ of genus curves so that the volume fraction on
the high-density region of the isodensity surface is equal to
\begin{equation}
   f = \frac{1}{\sqrt{2\pi}} \int_\nu^\infty e^{-t^2/2} dt.
\end{equation}
Adopting this method, the lognormal model and random-Gaussian model
would not be distinguished from each other (Coles \& Jones
1991). Since we intend to distinguish them so as to see whether or not
the lognormal model can fit the genus number, we adopt the
straightforward definition $\delta = \nu \sigma$ of the density
threshold throughout this paper.

To obtain the normalized genus curve $G(\nu)/G(0)$, we follow the
method developed in Matsubara \& Suto (1995). In practice, we first
compute $G(\nu)$ at 51 bins (in equal interval) for $-3 \leq \nu \leq
3$. Then we estimate the amplitude of $G(0)$ by $\chi^2$-fitting
the 7 data points around $\nu = 0$ to the lognormal formula
(\ref{eq:logn}) so that thus computed value of $G(0)$ is less affected
by the statistical fluctuation at one data point. The result is almost
insensitive to which fitting formula we use in estimating
$G(0)$.

The PDFs, $P(\nu)$, which have been similarly computed with 50
bootstrap resampling errors, and normalized genus curves, $G(\nu)/G(0)$, are
plotted in Figures 1 to 3 for power-law models with $n=-1$, $0$, and
$1$, respectively.
We select three different sets of the expansion factor $a$ ($=1$ at
the initial epoch) and the filtering length $R_f$ for each model so
that the resulting $\sigma(R_f)$ covers from weakly to fully nonlinear
regimes. The upper panels show the PDF and lower panels show the genus
curves.  The results of random Gaussian distribution,
(\ref{eq:pdflin}) and (\ref{eq:lin}), are plotted in dotted curves,
those of second-order perturbation theory, (\ref{eq:pdfsecond}) and
(\ref{eq:second}) with (\ref{eq:powerstu}) in dashed curves, lognormal
formulae, (\ref{eq:pdfln}) and (\ref{eq:logn}), in solid curves,
chi-square formulae, (\ref{onepoint}) and (\ref{chisquare}), in
dot-dashed curves. Symbols indicate the results of $N$-body
simulations. The curves for random Gaussian and second-order
perturbation theory using Edgeworth series are not guaranteed that the
genus for the negative density vanish. We have forced these
theoretical curves to be zero in the relevant regions in the plot.

The comparison of genus curves between the $N$-body results and the
theories based on the Edgeworth series are described in Matsubara \&
Suto (1995) in detail, so we do not repeat the detailed argument here.
In short, these two curves agree well for $-0.2 \lsim \nu\sigma
\lsim 0.4$ where perturbation theory is expected to be valid, but the
extrapolation of the second-order formula beyond this regime does
not work.

As for PDFs, the lognormal model fits the simulation results fairly
well from weakly to fully nonlinear regimes.  The degree of agreement,
however, depends on the initial power spectrum.  It works best for
$n=0$ model. Note that we use the Gaussian window. Usually, the
lognormal model is applied to the count-in-cells analysis which
corresponds to the top-hat window. For the top-hat window, $n=-1$
model fits the lognormal model better than $n=0$ model (Bernardeau \&
Kofman 1995).  Other non-Gaussian curves including chi-square
distribution fit simulation data well on weakly nonlinear regimes,
while they deviate considerably from simulation results on fully
nonlinear regimes.

For the genus, the lognormal model is also the best among the
non-Gaussian distributions considered here. The degree of agreement
between the lognormal model and simulation results also depends on the
initial power spectrum.  The fitting is the best for $n=-1$ model.
For other models, the lognormal model does not fit well in positive-threshold
 regions.  For all the simulation results, the agreement of the
lognormal model happen to be better than weakly nonlinear formula for
low-density regions. This may be partly because the weakly nonlinear
formula cannot naturally take into account the positivity of the
density, but the lognormal model achieves it by construction.  On the
other hand, the chi-square model does not agree well with any
simulation results except for the initial linear regime, even if it
assures the positivity of the density.

In Figure 4 are plotted the PDFs and normalized genus curves for LCDM
model for an example of a realistic cosmological scenario.
The smoothing length $R$ is $4\himpc$. If galaxies trace mass, $a=6$
corresponds to the present epoch ($z=0$). Thus $a=4$ and $5$
correspond to $z=0.5$ and $0.2$, respectively. We find our simulation
results of LCDM model and the lognormal model for Gaussian smoothed
PDFs agrees fairly well except for some differences of the peak height.
See also Ueda \& Yokoyama (1995) for top-hat smoothed PDF.  For the
normalized genus curves, the lognormal model fits the simulation
results well.

\section{Conclusions and Discussion} \label{sec:concl}

In the present paper, we have derived  theoretical prediction of the
genus statistics for some non-Gaussian distributions. We have generally
calculated the expectation value of genus number for non-Gaussian
fields that are given by  nontrivial functions of Gaussian random
fields. Two specific fields of this category, the lognormal distribution
and the chi-square distribution were investigated in detail.

As is seen in the figures the one-point PDF of the smoothed density
field is fitted by the lognormal model fairly well for all the
simulated models adopted here, {\it i.e.}, power-low models with
$n=-1,0,1$ and the LCDM model.  The lognormal formula for the genus
curve also works well to fit all the simulation data in the low
density regions.  But considerable deviation is observed in the
positive-threshold regions in the nonlinear regime of the power-law
models with $n=0$ and 1.  On the other hand, the formula fits the
entire regions of $n=-1$ and the LCDM models fairly well.  Thus the
genus statistics are more appropriate than one-point PDF to distinguish
between various initial power spectra for Gaussian smoothed field.
This is because the former depends on the spatial derivatives of
density field, too.

Several arguments exist to explain the validity of the lognormal model
as a statistical distribution in nonlinear regimes.  Coles \& Jones
(1991) argues that the lognormal distribution is obtained from the
continuity equation in the nonlinear regime but with linear or
Gaussian velocity fluctuations, so that it may be adopted as a model
to describe statistics in the weakly nonlinear regime.  On the other
hand, Bernardeau \& Kofman (1995) claims its successful fit to the PDF
of CDM type simulations is just a coincidence due to the particular
shape of the CDM power spectrum based on the top-hat smoothing.
Since
LCDM power spectrum has a similar structure to $n=-1$ power law in the
interested scales, our results of the genus curves are consistent with
these arguments.

Observationally, the negative binomial distribution, which has the
same one-point PDF as the chi-square distribution, has been shown to
fit the counts-in-cells of various redshift samples well (Gazta\~naga
\& Yokoyama 1993; Bouchet et al.\ 1993).  However, our results
indicate that chi-square formulae reproduce neither one-point PDF nor
genus curve of the simulations.  Does that imply these $N$-body
simulations have nothing to do with the real universe?  Not
necessarily, because here we are extracting information on relatively
small length scales $R\sim 4h^{-1}$Mpc using all the $64^3$ particles
in the simulations, while the observational counts analysis has been
performed on larger scales using volume-limited samples with much
smaller number density of galaxies.  The effect of sparse sampling in
the counts analysis has been analyzed by Ueda \& Yokoyama (1995) for
the LCDM model and it has been shown that the chi-square or the
negative binomial PDF does not fit the data well if we use all the
particles in the simulation but it fits well if we use
sparsely-sampled data with a similar number density to that of
volume-limited samples currently available.

Thus our results do not rule out the LCDM model.  On the contrary, it
remains one of the most promising models of our universe
(e.g., Suto 1993). To test the LCDM model further, we can use the
genus curve by examining if it is fitted by the lognormal formula.
Although the presently available redshift data are not
statistically significant
enough (e.g., Vogeley et al.\ 1994) to extract a specific
conclusion, we can reasonably expect the statistical significance of
observed genus curve will improve rapidly in the near future.

Now that we have obtained a number of theoretical formulae for the
genus curve we can make use of it not only to test the Gaussianity of
the primordial fluctuations but also to discriminate between various
models of structure formation.

\vskip 2cm

\centerline {\bf ACKNOWLEDGMENTS}

We are grateful to Yasushi Suto for providing us his $N$-body
simulation data and helping us to plot the genus curves from
simulation data and to David Weinberg for providing us the routines to
compute genus curve from numerical data.  T.~M.~gratefully
acknowledges the fellowship from the Japan Society of Promotion of
Science.  This research was supported in part by the Grants-in-Aid by
the Ministry of Education, Science and Culture of Japan (No.~0042).

\newpage

\centerline {\bf REFERENCES}
\bigskip

\def\pp{\par\parshape 2 0truecm 16.5truecm 1truecm 15.5truecm\noindent}
\def\apjpap#1;#2;#3;#4; {\pp#1, #2, #3, #4.}
\def\apjbook#1;#2;#3;#4; {\pp#1, #2 (#3: #4).}
\def\apjproc#1;#2;#3;#4;#5;#6; {\pp#1, {\sl #2} #3, (#4: #5), #6.}
\def\apjppt#1;#2; {\pp#1, #2.}

\apjbook Adler,~R.~J. 1981;The Geometry of Random
Fields;Chichester;Wiley;
\apjpap Bardeen,~J.~M., Bond,~J.~R., Kaiser,~N. \& Szalay,~A.~S. 1986;
ApJ;304;15;
\apjpap Bernardeau,~F. 1994;ApJ;433;1;
\apjpap Bernardeau,~F. \& Kofman,~L. 1995;ApJ;443;479;
\apjpap Bouchet,~F.~R., Juszkiewicz,~R., Colombi,~S. \& Pellat,~R.
 1992;ApJ;394;L5;
\apjpap Bouchet,~F.~R., Strauss,~M.~A., Davis,~M., Fisher,~K.~B.,
Yahil,~A. \& Huchra,~J. 1993;ApJ;417;36;
\apjpap Carruthers, P. 1991;ApJ;380;24;
\apjpap Coles,~P. \& Jones,~B. 1991;MNRAS;248;1;
\apjpap Doroshkevich,~A.~G. 1970;Astrophysics;6;320 (transl. from
Astrofizika, 6, 581);
\apjpap Fry,~J.~N. 1986;ApJ;306;358;
\apjpap Gazta\~naga, E \& Yokoyama, J. 1993;ApJ;403;450;
\apjpap Gott,~J.~R., Melott,~A.~L. \& Dickinson,~M. 1986;ApJ;306;341;
\apjpap Gott,~J.~R., Weinberg,~D.~H. \& Melott,~A.~L. 1987;ApJ;319;1;
\apjpap Gott,~J.~R., Miller,~J., Thuan,~T.~X., Schneider,~S.~E.,
Weinberg,~D.~H., Gammie,~C., Polk,~K., Vogeley,~M., Jeffrey,~S.,
Bhavsar,~S.~P., Melott,~A.~L., Giovanelli,~R., Haynes,~M.~P.,
Tully,~R.~B. \& Hamilton,~A.~J.~S. 1989;ApJ;340;625;
\apjpap Groth,E.J. \& Peebles,P.J.E. 1977;ApJ;217;385;
\apjpap Hamilton, A.~J.~S. 1985;ApJ;292;L35;
\apjpap Hamilton,~A.~J.~S. 1988;PASP;100;1343;
\apjpap Hamilton,~A.~J.~S., Gott,~J.~R. \& Weinberg,~D. 1986;
ApJ;309;1;
\apjpap Hernquist,~L., Bouchet,~F.~R., \& Suto,~Y. 1991;ApJS;75;231;
\apjpap Hubble,~E. 1934;ApJ;79;8;
\apjpap Juszkiewicz,~R., Weinberg,~D.~H., Amsterdamski,~P.,
Chodorowski,~M. \& Bouchet,~F. 1995;ApJ;442;39;
\apjpap Kofman,~L.~A., Bertschinger,~E., Gelb,~M.~J., Nusser,~A. \&
Dekel,~A. 1994;ApJ;420;44;
\apjpap Ling,~E.~N., Frenk,~C.~S. \&
Barrow,~J.~D. 1986;MNRAS;223;21{\sc p};
\apjpap {\L}okas,~E.~L., Juszkiewicz,~R., Weinberg,~D.~H. \&
Bouchet,~F.~R. 1995;MNRAS;274;730;
\apjpap Matsubara,~T. 1994;ApJ;434;L43;
\apjppt Matsubara,~T. 1995;UTAP-210/95 preprint;
\apjpap Matsubara,~T. \& Suto,~Y. 1994;ApJ;420;497;
\apjppt Matsubara,~T. \& Suto,~Y. 1995;ApJ, submitted;
\apjpap Melott,~A.~L., Weinberg,~D.~H. \& Gott,~J.~R. 1988;ApJ;328;50;
\apjpap Moore,~B., Frenk,~C.~S., Weinberg,~D.~H., Saunders,~W.,
Lawrence,~A., Ellis,~R.~S., Kaiser,~N., Efstathiou,~G. \&
Rowan-Robinson,~M. 1992;MNRAS;256;477;
\apjpap Okun,~B.~L. 1990;J.~Stat.~Phys.;59;523;
\apjpap Park,~C. \& Gott,~J.~R. 1991;ApJ;378;457;
\apjpap Park,~C., Gott,~J.~R. \& da Costa,~L.~N. 1992;ApJ;392;L51;
\apjpap Peebles, P.~J.~E. \& Groth, E.~J. 1975;ApJ;196;1;
\apjpap Rhoads,~J.~E., Gott,~J.~R. \& Postman,~M. 1994;ApJ;421;1;
\apjpap Suginohara,~T. \& Suto,~Y. 1991;PASJ;43;L17;
\apjpap Suginohara,~T. , Suto,~Y., Bouchet,~F.~R. \&
Hernquist,~L. 1991;ApJS;75;631;
\apjpap Suto, Y. 1993;Prog.~Theor.~Phys.;90;1173;
\apjpap Suto, Y. \& Matsubara, T. 1994;ApJ;420;504;
\apjpap Totsuji, H. \& Kihara, T. 1969;Publ. Astron. Soc. Japan.;21;221;
\apjppt Ueda,~H. \& Yokoyama,~J. 1995;YITP/U-94-27 preprint,
astro-ph/9501041;
\apjpap Vogeley,~M.~S., Park.~C., Geller,~M.~J., Huchra,~J.~P. \&
Gott,~J.~R. 1994;ApJ;420;525;
\apjpap Weinberg,~D.~H. 1988;PASP;100;1373;
\apjpap Weinberg,~D.~H. \& Cole,~S. 1992;MNRAS;259;652;
\apjpap White, S.~D.~M. 1979;MNRAS;186;145;

\vskip 1.5cm
\centerline{\bf FIGURE CAPTIONS}

\bigskip
\begin{description}

\item[Figure 1]{The PDFs (upper panels) and normalized genus
curves (lower panels) from the $N$-body simulation data for $n = - 1$
power-law model ($\Omega_0 = 1$, $\lambda_0 = 0$) are plotted by open
circles. Three different sets of the expansion factor $a$ ($=1$ at the
initial epoch) are selected and the Gaussian window function with the
filtering length $R_f = L/25$ is used. The values of the expansion
factor and the resulting variance $\sigma$ are indicated in the
figure. The theoretical prediction for the random Gaussian field is
plotted by dotted lines, lognormal model by solid lines, chi-square
model by dot-dashed lines, second-order perturbation theory by dashed
lines. \label{fig1}}
\item[Figure 2]{Same as Fig.~\protect\ref{fig1} for $n = 0$
power-law model. \label{fig2}}
\item[Figure 3]{Same as Fig.~\protect\ref{fig1} for $n = 1$
power-law model. \label{fig3}}
\item[Figure 4]{Same as Fig.~\protect\ref{fig1} for low-density cold
dark matter model ($\Omega_0 = 0.2$, $\lambda_0 = 0.8$, $h = 1.0$) The
adopted Gaussian filtering length $R_f$ corresponds to $4 \himpc$
(comoving). The top-hat smoothed rms mass fluctuation at $a = 6$ is
unity at $8 \himpc$. \label{fig4}}
\end{description}
\end{document}